\documentclass[final,1p,times]{elsarticle}

\usepackage{graphicx}
\usepackage{amssymb}

\journal{Physics A}

\begin{document}
\begin{frontmatter}

\title{Collaborative filtering with diffusion-based similarity on tripartite graphs}

\author[inst1]{Ming-Sheng Shang}
\author[inst2]{Zi-Ke Zhang}
\author[inst2,inst3]{Tao Zhou}
\author[inst1,inst2]{Yi-Cheng Zhang}

\address[inst1]{Web Sciences Center, University of
Electronic Science and Technology of China, 610054 Chengdu, P. R.
China}
\address[inst2]{Department of Physics, University of Fribourg, Chemin du Mus\'ee
3, 1700 Fribourg, Switzerland}
\address[inst3]{Department of Modern Physics, University of Science and Technology of China, Hefei
Anhui 230026, P. R. China}

\begin{abstract}
Collaborative tags are playing more and more important role for the
organization of information systems. In this paper, we study a
personalized recommendation model making use of the ternary
relations among users, objects and tags. We propose a measure of
user similarity based on his preference and tagging information. Two
kinds of similarities between users are calculated by using a
diffusion-based process, which are then integrated for
recommendation. We test the proposed method in a standard
collaborative filtering framework with three metrics: ranking score,
Recall and Precision, and demonstrate that it performs better than
the commonly used cosine similarity.
\end{abstract}


\end{frontmatter}

\section{Introduction}
\label{Introduction} With the rapid growth of the Internet
\cite{ZhangGQ2008} and the World-Wide-Web \cite{Broder2000}, a huge
amount of data and resource is created and available for the public.
This, however, may result in the problem of \emph{information
overload}: we face an excess amount of information, and are unable
to find the relevant objects. In consequence, it is vital to study
how to automatically extract the hidden information and make
personalized recommendations. There have been a number of
significant works trying to solve this problem. A landmark is the
use of search engine \cite{Brin1998,Kleinberg1999}. However, a
search engine could only find the relevant web pages according to
the input keywords and return the same results regardless of users'
habits and tastes. An alternative is the use of the recommender
system \cite{Resnick1997,Belkin2000}, which is, essentially, an
information filtering technique that attempts to present information
likely of interest to the user. Due to its significance for economy
and society, the design of efficient recommendation algorithms has
become a common focus of branches of science (see the review
articles \cite{Adomavicius2005,Herlocker2004} and the references
therein).

Typically, a recommender system compares the user's profile to some
reference characteristics, and seeks to predict the `rating' that a
user would give to an object he had not yet considered. The
mainstream of recommendation algorithms can be divided into two
categories \cite{Adomavicius2005}: (i) the content-based methods in
which the recommended objects are similar to those preferred by the
target user in the past; (ii) the collaborative filtering (CF) in
which the recommended objects are popular among the users who have
similar preferences with the target user. Thus far, CF is the most
successful method underlying recommender systems. Over the last
decade many algorithms under the CF framework have been proposed,
including similarity based approaches
\cite{Adomavicius2005,Herlocker2004}, relevance models
\cite{Wang2008}, matrix factorization techniques \cite{Weimer2008},
iterative self-consistent refinement \cite{Ren2008}, and so on.

A fundamental assumption of CF method is that, in a social network,
those who agreed in the past tend to agree again in the future. The
most commonly used algorithms in CF is a neighborhood-based
approach, which works by first computing similarities between all
pairs of users, and then to predict by integrating ratings of
neighbors (i.e., those who having high similarities to the target
user) of the target user. Algorithms within this family differ in
the definition of similarity, formulation of neighborhoods and the
computation of predictions. There are two main algorithmic
techniques \cite{Adomavicius2005}: user-based and object-based,
which are mathematically equivalent by interchanging the roles of
user and object; in this paper, we will only consider the user-based
technique. The most crucial step for collaborative filtering is to
find a particular user's neighborhood with similar taste or interest
and quantify the strength of similarity
\cite{Adomavicius2005,Herlocker2004,Wang2008,Breese1998}. Various
kinds of methods have been proposed on this issue (see Refs.
\cite{Lee2008,Liu2009,Sun2009} for some recent works, to name just a
few), among which the {\it cosine similarity} \cite{Salton1983} and
the \emph{Jaccard index} \cite{Jaccard1901} are the most commonly
used measurements.

Most of previous studies only consider the ratings given to the
object, while neglect the content information. A possible reason is
that the content information is hard to automatically extracted out,
and how to properly make use of such information is not known well.
Very recently, collaborative tagging systems have been introduced
into the studies of recommender systems \cite{Tso2008,Zhang2009}. In
collaborative tagging systems (CTSes), users are allowed to freely
assign tags to their collections, which can both express users'
personalized preferences and describe the objects' contents. In Ref.
\cite{Tso2008}, tags are incorporated to the standard CF algorithm
by reducing the three-dimensional correlations to three
two-dimensional correlations and then applying a fusion method to
re-associate these correlation. In Ref. \cite{Zhang2009}, a
recommendation algorithm via integrated diffusion on user-object-tag
tripartite graphs is studied. In this paper, we propose a
collaborative filtering algorithm based on a new measure of user
similarity which integrates user preferences of both collected
objects and used tags. We evaluate our method on a benchmark data
set, \emph{MovieLens}. Experimental results demonstrate that our
method can outperform the standard CF based on cosine and Jaccard
indices.

\begin{figure}
\begin{center}
       \center \includegraphics[width=13cm]{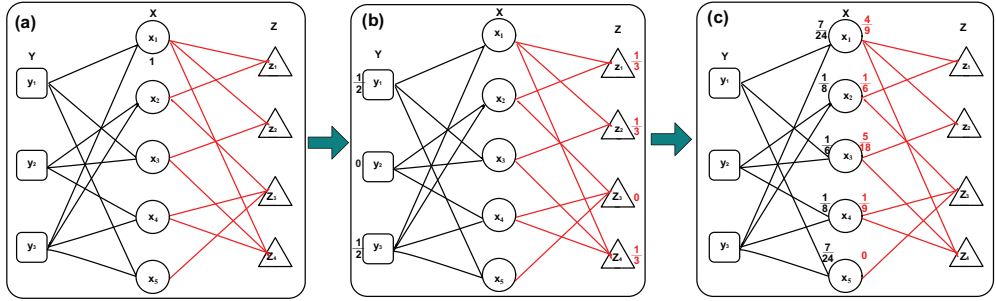}
\caption{(Color online) Illustration of the diffusion-based
similarity on a tripartite graph. Plot (a) shows the initial
condition where the target user $x_1$ is assigned a unit of
resource; plot (b) describes the result after the first-step
diffusion, during which the resource is transferred equally from
user $x_1$ to objects $y\in Y$ and tags $z\in Z$; eventually, the
resources flow back to users, and we show the result in plot (c).
The values marked beside nodes in black and red are respectively
denote the amounts of resource in the user-object and user-tag
diffusions.} \label{fig1}
\end{center}
\end{figure}

\begin{table}
\renewcommand{\arraystretch}{1.3}
\caption{The best algorithmic performance for ranking score, Recall
and Precision. DS and CS are abbreviations of diffusion similarity
and cosine similarity. The numbers in the brackets are the
corresponding optimal values of $\lambda$. The results reported here
are consistent with what shown in Figs. 2-4. Note that for all three
metrics, the diffusion similarity performs much better than the
cosine similarity.}
\begin{tabular}{cccccc}
\hline \hline
 & $\langle RankS\rangle$ & $R$($L=10$) & $R$($L=20$) & $P$($L=10$) & $P$($L=20$) \\
\hline
DS & 0.19943(0.74) & 0.08469(0.62) & 0.12333(0.62) & 0.00931(0.74) & 0.00698 (0.80) \\
CS & 0.21973(0.62) & 0.00626(1.00) & 0.01071(1.00) & 0.00095(1.00) & 0.00082 (0.00) \\
\hline \hline
\end{tabular}
\end{table}

\section{Method}
\label{Method} In the system, there are three kinds of elements,
users, objects and tags. Each user has collected some objects and
described them with tags. Let $U$ be a set of $m$ users, $O$ be a
set of $n$ objects, and $T$ be a set of $r$ tags. The relationships
among the three sets can be described by a tripartite graph. In this
paper, we are interested in the similarities among users, and thus
can reduce this tripartite graph into two pair correlations: {\it
user-object} and {\it user-tag}, which can be described by two
adjacent matrices, $A$ and $A'$, respectively. If user $u$ has
collected object $\alpha$, we set $a_{u\alpha}=1$, otherwise
$a_{u\alpha}$ = 0. Analogously, we set $a'_{us}=1$ if $u$ has used
the tag $s$, and $a'_{us}=0$ otherwise.

\subsection{Diffusion-Based Similarity}
We use a diffusion process to obtain similarities between users
\cite{Ou2007,Zhou2007}. The basic idea is shown in Fig. \ref{fig1}.
Considering the user-object bipartite graph, and assume that a unit
of resource (e.g. recommender power) is associated with the target
user $v$, which will be distributed  to other users, such that each
user gets a specific percentage. At the first step, the user $v$
distributes the resource equally to all the objects he has
collected. After this step, the resource that object $\alpha$ gets
from $v$ reads
\begin{equation}
  r_{\alpha v}=\frac{a_{v\alpha}}{k(v)},
\end{equation}
where $k(v)$ is the {\it degree} of $v$ in the user-object bipartite
graph. Then, at the second step, each object distributes it's
resource equally to all the users having collected it. Thus,
resource that $u$ gets from $v$, which we define as {\it similarity}
between $u$ and $v$ with $v$ the target user (note that, this
similarity measure is asymmetry), is:
\begin{equation}
\label{eq1}
{s_{uv}} = \sum_{\alpha \in O} {\frac{a_{u\alpha}\cdot r_{\alpha v}}{k(\alpha)}}
         =  \frac{1}{k(v)}\sum_{\alpha\in O}{\frac{a_{u\alpha}
         a_{v\alpha}}{k(\alpha)}},
\end{equation}
where $k(\alpha)$ is the degree of object $\alpha$ in the
user-object bipartite graph, and $O$ is the set of objects.

Analogously, considering the diffusion on the user-tag bipartite
graph. Suppose that a unit of resource is initially located on the
target user $v$, which will be equally distributed to all tags he
has used, and then each tag redistributes the received resource to
all its neighboring users. Thus, we obtain tag-based similarity
between user $u$ and $v$ (with $v$ the target), as
\begin{equation}
\label{eq2} {s'_{uv}} =  \frac{1}{k'(v)}\sum_{t\in T}{\frac{a'_{ut}
a'_{vt}}{k'(t)}} ,
\end{equation}
where $k'(t)$ and $k'(v)$ are respectively the degrees of tag $t$
and user $v$ in the user-tag bipartite graph, and $T$ is the set of
tags.

\subsection{Recommendation with Integrated Similarity}
Tso \emph{et al.} \cite{Tso2008} and Zhang \emph{et al.}
\cite{Zhang2009} have recently demonstrated the significance of
making use of the CTSes to improve the accuracy of recommendations.
Motivated by those results, we plan to integrate the above two
diffusion-based similarities to get better recommendations. As a
start point, in this paper, we adopt the simplest way, that is, to
combine $s_{uv}$ and $s'_{uv}$ linearly:
\begin{equation}
s_{uv}^*= \lambda s_{uv}+(1-\lambda)s'_{uv},
\end{equation}
where $\lambda \in [0,1]$ is a tunable parameter. For cosine and
Jaccard indices, we also firstly get the similarities respectively
based on user-object and user-tag bipartite graphs, and then
integrate them in a linear way as shown in Eq. (4). Since the
Jaccrad index performs almost the same as the cosine similarity,
this paper only reports the numerical results on cosine similarity.

Next, we apply the standard collaborative filtering for
recommendation \cite{Herlocker2004}. Given a target user $v$ and an
uncollected object $\alpha$, the preference of $v$ on $\alpha$ is:
\begin{equation}
\label{eq4} p_{v\alpha}=\sum_{u \neq v} s^{*}_{uv}a_{u\alpha}.
\end{equation}
We then sort all objects that user $v$ has not collected in the
descending order of their scores, and the top $L$ objects will be
recommended to $v$.

\begin{figure}
    \begin{center}
       \center \includegraphics[width=10cm]{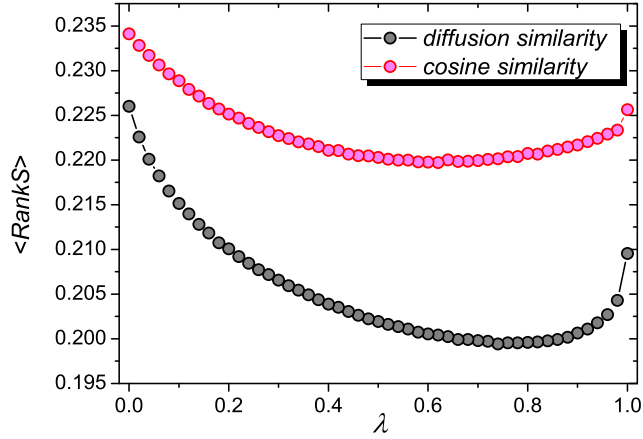}
        \caption{(Color online) $\langle RankS \rangle$ versus $\lambda$. The results reported here are
                 averaged over 5 independent runs, each of which corresponds to a
                 random division of training set and testing set. $\lambda$=0 and $\lambda$=1 correspond to the
                 cases for pure user-tag and user-object diffusions, respectively. The two curves are
                 corresponding to diffusion-based similarity (lower black) and cosine similarity (upper red), respectively.
                 The smaller value indicates the higher accuracy of recommendation algorithm.}
    \label{fig2}
    \end{center}
\end{figure}

\section{Numerical Results}
\label{Experiment}

\subsection{Data Set}
In this paper, we use a benchmark data, \emph{MovieLens}
(http://www.grouplens.org), to evaluate our proposed algorithm.
\emph{MovieLens} is a movie rating system, where each user votes
movies in five discrete ratings 1-5 and a tagging function was added
since January 2006. With the help of collaborative tags, users can
look into the pool of movies that are assigned with the same tag. We
here only consider the objects and tags having been collected and
used by at least two users, and the users who have collected and
used at least one object and one tag. The sampling data consists of
3710 users, 5724 objects and 5228 tags, with 53091 user-object and
33065 user-tag relations. To test the algorithmic performance, in
each run, the data set is randomly divided into two parts: the
training set contains 90\% of entries, and the remaining 10\%
constitutes the testing set.

\begin{figure}
    \begin{center}
       \center \includegraphics[width=10cm]{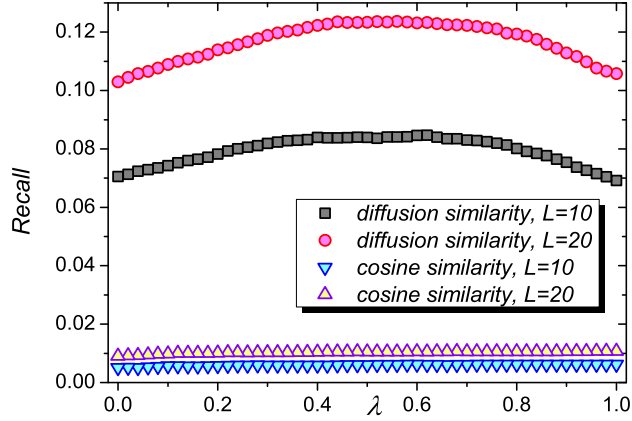}
        \caption{(Color online) \emph{Recall} versus $\lambda$. The results reported here are
                 averaged over 5 independent runs, each of which corresponds to a
                 random division of training set and testing set. $\lambda$=0 and $\lambda$=1 correspond to the
                 cases for pure user-tag and user-object diffusions, respectively.
                 The higher value indicates the higher accuracy of recommendation algorithm.}
    \label{fig3}
    \end{center}
\end{figure}

\subsection{Metrics for Algorithmic Performance}
We employ three metrics, \emph{ranking score} (RankS)
\cite{Zhou2007}, \emph{Recall} \cite{Herlocker2004} and
\emph{Precision} \cite{Herlocker2004}, to investigate the
performance of the proposed algorithm, the former one takes into
account the whole rank of objects and the latter two concern only
the objects with the highest scores, i.e., the recommended objects.
\begin{enumerate}
\item \emph{RankS}.--
\emph{RankS} describes the position of the uncollected objects. That
is, if the edge $u-\alpha$ ($u$ is a user and $\alpha$ is an object)
is in the testing set, we calculate the position of $\alpha$ of all
the uncollected objects of $u$, and denote it as $r_{u\alpha}$. For
example, if there are 100 uncollected objects for $u_i$ and $\alpha$
is put in the third, then $r_{u\alpha}=0.03$. Since the objects in
the testing set are actually collected by users, smaller
$r_{u\alpha}$ is favored. The average of $r_{u\alpha}$ over all
user-object pairs in the testing set defines the average ranking
score, as:
\begin{equation}
  \langle RankS \rangle=\frac{1}{N_p}\sum_{(u,\alpha) \in E^T}r_{u\alpha},
\end{equation}
where $E^T$ is the set of user-object pairs in the testing set,
$N_p$ is the number of elements in $E^T$. Clearly, the smaller the
$\langle RankS \rangle$, the higher the accuracy.
\item \emph{Recall}.---
\emph{Recall} is the ratio of relevant objects in the recommendation
list to the total number of the relevant objects (i.e., the total
number of user-object pairs in the testing set). It reads
\begin{equation}
  \emph{R}=\frac{1}{N_p}\sum_{u\in U}N^u_r,
\end{equation}
where $N^u_r$ is the number of recommended objects for user $u$ that
are indeed in the testing set. $R$ depends on the length of
recommendation list, and the larger the $R$ the higher the accuracy.
\item \emph{Precision}.---
\emph{Precision} is the ratio of relevant objects in the
recommendation list to the total number of the recommended objects.
It reads
\begin{equation}
  \emph{P}=\frac{1}{mL}\sum_{u \in U}N^u_r,
\end{equation}
$P$ depends on the length of recommendation list, and the larger the
$P$ the higher the accuracy.
\end{enumerate}

\subsection{Experimental Results}
Figure 2 shows the $\langle RankS \rangle$ of the two kinds of
similarities, diffusion-based similarity and cosine similarity, as a
function of the parameter $\lambda$. It can be seen that both two
kinds of similarities can get benefit by making use of tag
information, namely can reach lower $\langle RankS \rangle$ with
proper $\lambda$. Comparing with the algorithm without tag
information, at the optimal values, the improvements for
diffusion-based similarity and cosine similarity are 4.83\% and
2.62\%, respectively. In addition, the diffusion-based similarity
performs better that the cosine similarity under the standard CF
framework.

Figure 3 reports \emph{Recall} as a function of $\lambda$. Since the
typical length for recommendation list is tens, our experimental
study focuses on the interval $L\in [10,100]$. To keep the figure
neat, we only show the results for $L=10$ and $L=20$, with $\lambda
\in [0,1]$. Different from the case of $\langle Ranks \rangle$, the
tag information does not contribute much to the \emph{Recall} for
cosine similarity, and it contributes some but not much to the
diffusion-based similarity. This may be caused by the data sparsity
for user-tag relations, which is known as a typical reason leading
to the ineffectiveness of CF. There are almost the same number of
objects and tags (5724 vs. 5228), but the number of user-tag
relations is 60\% less than that of user-object relations. That is
to say, the density of data may also be a crucial ingredient for
recommendation of collaborative tagging systems, and only if the tag
information is rich, one can get benefit from it. Similar results
for Precision are presented in Fig. 4. In Table 1, we summarize the
optimal values for the three accuracy metrics, which again
demonstrate that the diffusion-based similarity could provide
remarkably better recommendations than the cosine similarity.

\begin{figure}
    \begin{center}
       \center \includegraphics[width=10cm]{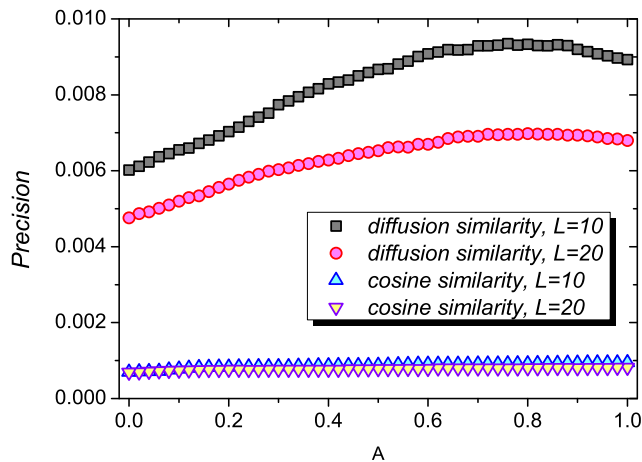}
        \caption{(Color online) \emph{Precision} versus $\lambda$. The results reported here are
                 averaged over 5 independent runs, each of which corresponds to a
                 random division of training set and testing set. $\lambda$=0 and $\lambda$=1 correspond to the
                 cases for pure user-tag and user-object diffusions, respectively.
                 The higher value indicates the higher accuracy of recommendation algorithm.}
    \label{fig4}
    \end{center}
\end{figure}

\section{Conclusions}

In this paper, we proposed an integrated diffusion-based similarity
with the help of tag information. Experimental results demonstrate
that the tag information can be used to improve the accuracy of
recommendations. In addition, the diffusion-based similarity works
much better than the cosine and Jaccard similarity. There are many
topology-based similarity indices, some of them are based on local
information, while others require global knowledge of network
structure (see References \cite{Liben-Nowell2007,Zhou2009}). Some of
them can not be easily extended to the bipartite graphs (e.g., Katz
index, average commute time, etc.) and the calculation of global
indices is very time consuming. Since the diffusion-based similarity
requires no more calculation than the cosine and Jaccard indices, we
believe it could find the application in real recommender systems.

The present algorithm depends on a free parameter $\lambda$. In the
case of $\lambda=1$, it degenerates to the algorithm not making use
of tag information at all. Therefore, to compare the $\lambda=1$
case with the optimal case, one could see how tag information can
help improving the algorithmic accuracy. An interesting result is
that the diffusion-based similarity can make better use of tag
information than the cosine and Jaccard indices. In addition, in
comparison to the results reported by Zhang \emph{et al.}
\cite{Zhang2009}, the present algorithm has much higher values of
Recall.

The collaborative tagging systems are playing more and more
important role in the Internet world, and we must be aware of their
significance. Experimental results in this paper strongly suggest
using the tag information to improve the quality of recommendations.
Indeed, we should encourage users to try to experience online
systems with tags, particularly for organizing personal interests.
Although in the beginning, users may assign each object with
arbitrary number of tags, previous researches have revealed that the
tag vocabulary will grow in a sub-linear way both in open
\cite{Cattuto20071} and canonical \cite{Zhang20092} systems. In
addition, in the statistically level, the number of tags associated
with each tagging action will converge to a certain value
\cite{Cattuto20071}.

This paper only provides a simple beginning for the design of
recommendation algorithms making use of tag information. There are
still many open issues remain for the further study. First, the more
in-depth understanding of the structure of collaborative tagging
systems would be helpful for generating better recommendations.
Second, since the tag information is considered to be a meaningful
accessory towards semantic relations for users and objects
\cite{Xu2006}, despite its sparsity problems, it should draw
potential yet promising relations for personalized recommendation
via community detection algorithms. Finally, this work only
considers the unweighted case for user-tag relations, however, a
user may assign different objects the same tag, making a weighted
relations between users and tags. Study of the weighted version may
give more insights and further improvements of recommender systems.

\section{Acknowledgments}
We acknowledge \emph{GroupLens Research Group} for \emph{MovieLens}
data (http://www.grouplens.org). This work is supported by China
Postdoctoral Science Foundation 20080431273, the 863 Project
2007AA01Z440, the Swiss National Science Foundation 200020-121848,
the Sino-Swiss Science and Technology Cooperation (SSSTC) Project EG
20-032009, and the National Natural Science Foundation of China
60973069. T.Z. acknowledges the National Natural Science Foundation
of China (Grant Nos. 10635040 and 60744003).

\end{document}